# Novel extraction algorithm for amplitude and phase measurement of ultrashort optical pulses via spectral phase interferometry

**Alessia Pasquazi,[1,\*] Marco Peccianti,[2] José Azaña [1], David J. Moss,[3,4] and Roberto Morandotti[1]**

[1]*INRS Énergie, Matériaux et Télécommunications, 1650 Blvd. Lionel Boulet, Varennes (Québec), J3X 1S2 Canada*
[2]*Institute for Complex Systems – CNR, UOS Montelibretti, Via dei Taurini 19, 00185 Roma, Italy.*
[3]*Centre for Ultra-high Bandwidth Devices for Optical Systems (CUDOS),and the Institute of Photonics and Optical Science (IPOS), School of Physics, University of Sydney, Sydney, NSW 2006,Australia*
[4]*David Moss Present Address: School of Electrical and Computer Engineering, RMIT University, Melbourne, Victoria, Australia 3001.*
[\*] *alessia.pasquazi@gmailcom*

## Summary

We present a novel extraction algorithm for spectral phase interferometry for direct field reconstruction (SPIDER) for the so-called X-SPIDER configuration. Our approach largely extends the measurable time windows of pulses without requiring any modification to the experimental X-SPIDER set-up.

# 1. Introduction

Phase-sensitive measurement techniques [1] for ultrafast (< 1ps) optical pulses with large spectral bandwidths, and over long temporal windows (>100ps) is becoming increasingly important, driven in no small part by the recent introduction of coherent optical communications [2-3]. In addition, recent progress in synthetizing arbitrary optical waveforms [4-6] has intensified the effort toward providing simple and practical metrological methods to measure complex pulses having large time-bandwidth products (TBPs). These developments are creating a growing and compelling need for ultrafast coherent optical pulse measurement techniques that can operate at milliwatt peak power levels and on timescales ranging from sub-picoseconds to nanoseconds. Previous reports of ultrafast optical signal measurements in integrated, CMOS compatible platforms include time-lens temporal imaging and waveguide-based frequency-resolved optical gating (FROG)[1,7-10]. These approaches transferred in the integrated domain two popular methods [1,11-12] for ultrafast pulse measurement; however, time-lens imaging is phase-insensitive while waveguide-based FROG methods require integrated long tunable delay lines - still an unsolved challenge for the full integration of the technique.

Sheared interferometry, introduced by Wamsley [13] in 1998, is one of the most widely used methods for characterizing ultrafast optical pulses due to its ability to recover the full complex (i.e. amplitude and phase) information of an ultra-fast optical waveform. Most of its strength is due to a direct and robust algorithm [13-15] that retrieves the pulse from the spectrum of two replicas of the same pulse, shifted (sheared) in frequency. The first and most popular implementation of sheared interferometry is named s*pectral phase interferometry for direct field reconstruction*, or SPIDER [1,13-14]. In the standard implementation of the SPIDER technique, the spectral shear between two replicas of the pulse under test (PUT) is obtained by nonlinearly mixing two delayed replicas of the PUT with a chirped pump pulse via a three wave mixing (TWM) process, as depicted in Fig. 1.

Each replica of the PUT must overlap with an almost monochromatic portion of the pump pulse in order to produce an idler with the same spectral shape of the PUT. As the two replicas mix with two different pump frequencies, the idler replicas result in two pulses that are sheared in frequency. The pump pulse can be a dispersed replica of the PUT itself making this method remarkably self-referenced. In many cases, however, it is possible to use an external, well-characterized pump pulse, thus improving the accuracy of the method. In this case, the technique is usually referred to cross –SPIDER or X-SPIDER[16-17]. In its many variants[1,13-20], SPIDER methods have proven to be well-suited for phase-sensitive characterization of ultra-fast laser pulses as short as 6fs, from the infrared to the ultraviolet.

Quite remarkably, the SPIDER method is fully compatible with designs that are amenable to waveguide implementation: recently [21], we reported a device capable of characterizing both the amplitude and phase of ultrafast optical pulses with the aid of a synchronized, incoherently related clock pulse. In this previous work, we redesigned the X-SPIDER approach in order to exploit degenerate four-wave mixing (DFWM) interactions to allow its implementation in centro-symmetric materials. We demonstrated that the X-SPIDER can be implemented in CMOS compatible integrated platforms, offering a simple phase sensitive metrological device on a chip. To address the demand for efficient methods to monitor optical data streams with high TBP in optical networks, we introduced a novel extraction algorithm method, that we termed *FLEA*: *Fresnel-limited extraction algorithm*. The FLEA dramatically improves the accuracy of any X-SPIDER device (based both on $3^{rd}$ order and more conventional $2^{nd}$ order nonlinearities) for optical pulses having very large TBPs: thanks to this approach, we measured [21] pulses with a frequency bandwidth of >1 THz stretched up to 100 ps pulsewidths, yielding a TBP of >100.

The FLEA removes a classical approximation in the X-SPIDER methods operating via nonlinear optical processes. In general, an X-SPIDER approach relies on the assumption of a pump pulse having a much larger chirp than the PUT. If this condition is violated, the two idler replicas obtained by the wave mixing process depicted in Fig. 1 are no more spectral sheared replicas of the PUT. Hence, the information retrieved with the usual algorithm will exhibit a significant error. For this reason SPIDER methods are usually restricted to application on pulses with low TBP. Indeed, in most cases, significantly increasing the pump chirp is not an available option as it is accompanied by a reduction in the signal-to-noise ratio of the reconstructed profile. Better results in terms of TBP have been obtained using variations of sheared interferometry that exploit spatial encoding, although at the expense of sacrificing simplicity, as well as the likelihood of achieving an integrated solution[22]. In any event, it is interesting to note that the nonlinear product of a standard SPIDER implementation always contains information on the phase of the PUT, and some retrieval strategies have been adopted for Gaussian pulses in the self-referenced case [20]. As detailed below, FLEA recognizes that the two idler replicas resulting in an X-SPIDER set-up are a *Fresnel Integral* of the PUT, and implements the reconstruction process according to this observation, significantly extending the operating regime (TBP) of existing X-SPIDER set-ups..

In this paper, we present the full detailed theory of FLEA for X-SPIDER measurements, and include detailed discussion of the inherent limits and tolerances of this phase-recovery approach. Our conclusions are validated through numerical examples on a standard test bench of pulses.

## 2. Parametric interaction and X-SPIDER

The PUT is represented in time and frequency by the temporal complex envelope of the electric field $e(t)$ and its spectrum $E(\omega)$, related to each other by the Fourier Transform:

$$e(t) = |e(t)| \exp[i\varphi_e(t)] = \int_{-\infty}^{\infty} E(\omega) \exp(-i\omega t) \frac{d\omega}{2\pi}$$
$$E(\omega) = |E(\omega)| \exp[i\varphi_E(\omega)] = \int_{-\infty}^{\infty} e(t) \exp(i\omega t) dt$$

(2.1)

As is typically done, we restrict our analysis to bandwidth-limited pulses, i.e. where the PUT vanishes outside of intervals given in time and frequency by $\Delta\tau_E$ and $\Delta\omega_E$, respectively. Both for TWM and DFWM, the nonlinear interaction involves three optical waves: a pump $p(t)$ that amplifies a signal $s(t)$ and generates an idler $i(t)$. As depicted in Fig. 1, the signal $s(t)$ consists of two delayed replicas of the PUT $e(t)$:

$$s(t) = \left[ e\left(t + \frac{\Delta t}{2}\right) + e\left(t - \frac{\Delta t}{2}\right) \right]$$

(2.2)

The pump pulse is chirped in order to produce a temporal quadratic phase curvature with an envelope that is approximately constant over the duration where nonlinear mixing with the signal occurs. Such a profile can be easily obtained by temporally stretching a transform limited pulse $p_o(t)$ with a flat-phase, smoothly varying spectrum $P_o(\omega)$, e.g. a spectrally bell-shaped or spectrally flat-top pulse over an interval $\Delta\omega_P$. After propagation through a predominantly first-order dispersive element with a total dispersion of $\phi_P = \beta_2 L$ (where $L$ is the propagation length and $\beta_2$ is the group velocity dispersion of the dispersive element), the pump can be expressed in the frequency domain as:

$$P(\omega) = P_o(\omega)\exp\left(-\frac{i\omega^2\phi_P}{2}\right) \qquad (2.3a)$$

In the temporal domain, the pump is the *Fresnel Integral* of its transform-limited version $p_o(t)$:

$$p(t) = \frac{1}{\sqrt{2i\pi\phi_P}}\exp\left(\frac{it^2}{2\phi_P}\right) * p_o(t) = \frac{1}{\sqrt{2i\pi\phi_P}}\exp\left(\frac{it^2}{2\phi_P}\right)\int_{-\infty}^{\infty} p_o(\tau)\exp\left(\frac{i\tau^2}{2\phi_P} - \frac{i\tau t}{\phi_P}\right)d\tau \qquad (2.3b)$$

Where the symbol * indicates the convolution operation. Under the assumption that $\phi_P(\Delta\omega_P)^2 >> 2\pi$, i.e. for a highly stretched pump, it is possible to use the Fraunhofer approximation of 2.3b:

$$p(t) \approx \frac{1}{\sqrt{2i\pi\phi_P}}\exp\left(\frac{it^2}{2\phi_P}\right)\int_{-\infty}^{\infty} p_o(\tau)\exp\left(-\frac{i\tau t}{\phi_P}\right)d\tau = \frac{1}{\sqrt{2i\pi\phi_P}}P_o\left(-\frac{t}{\phi_P}\right)\exp\left(\frac{it^2}{2\phi_P}\right) \qquad (2.3c)$$

In this case the temporal amplitude of the stretched pulse follows the smoothly varying spectral shape $P_o(\omega)$, e.g. with a bell shape or a nearly flat-top profile. The shape of the idler is dependent on the specific nonlinear process involved; in particular, considering valid the approximation 2.3c, we have:

$$i(t) \propto s(t)p(t) \propto s(t)P_o\left(-\frac{t}{\phi_P}\right)\exp\left(\frac{it^2}{2\phi_P}\right) \approx s(t)\exp\left(\frac{it^2}{2\phi_P}\right) \qquad (2.4a)$$

$$i(t)^* \propto s(t)p(t)^* \propto s(t)P_o\left(-\frac{t}{\phi_P}\right)\exp\left(-\frac{it^2}{2\phi_P}\right) \approx s(t)\exp\left(-\frac{it^2}{2\phi_P}\right) \qquad (2.4b)$$

$$i(t)^* \propto s(t)\left[p(t)^*\right]^2 \propto s(t)\left[P_o\left(-\frac{t}{\phi_P}\right)\right]^2\exp\left(-\frac{it^2}{\phi_P}\right) \approx s(t)\exp\left(-\frac{it^2}{\phi_P}\right) \qquad (2.4c)$$

(2.4a-c) reports the idler for the cases TWM sum frequency generation (SFG), TWM difference frequency generation (DFG) and DFWM, respectively. The last approximations in the relations above are valid when the pump temporal amplitude or the *square of the pump temporal amplitude* for the TWM and DFWM cases respectively is almost constant along the temporal window $\Delta\tau_S$ occupied by the signal:

$$\Delta\tau_S = \Delta t + \Delta\tau_E \leq \Delta\tau_{PN} \qquad (2.5a)$$

$\Delta\tau_{PN}$ is the temporal window defined by the pump in the nonlinear interaction and it is dependent on the specific shape taken into consideration. Table 1 summarizes the results for Gaussian and flat top pulses in the different nonlinear mixing cases.

As outlined in Fig.1, the idler is collected by a spectrometer, and so the experimental quantity of interest is the Power Spectral Density (PSD) $|I(\omega)|^2$ of the idler, for the cases above:

$$|I(\omega)|^2 \propto \left| S(\omega) * \exp\left(-\frac{i\omega^2 \phi_P}{2}\right) \right|^2 \qquad \text{TWM-SFG} \qquad (2.6a)$$

$$|I(\omega)|^2 \propto \left| S(\omega) * \exp\left(\frac{i\omega^2 \phi_P}{2}\right) \right|^2 \qquad \text{TWM-DFG} \qquad (2.6b)$$

$$|I(\omega)|^2 \propto \left| S(\omega) * \exp\left(\frac{i\omega^2 \phi_P}{4}\right) \right|^2 \qquad \text{DFWM} \qquad (2.6c)$$

For simplicity, in what follows, we will refer explicitly only to the case of SFG. As clear from the relations above, these results can be easily extended to the DFG case by changing the sign of the pump chirp, and to the DFWM case by changing the sign and dividing by 2 the pump chirp. Table 1 summarizes the results in the three nonlinear cases. Focusing on the SFG case, it is useful to define the following function:

$$F_s(\omega) = S(\omega) \exp\left(-\frac{i\omega^2 \phi_P}{2}\right) \qquad (2.7a)$$

The latter simply represents the output spectrum of the signal after an equivalent propagation in a first order dispersive system with the total dispersion of the pump $\phi_P$, i.e. its Fourier Transform is the *Fresnel Integral* of the signal:

$$f_s(t) = \frac{1}{\sqrt{2i\pi\phi_P}} \exp\left(\frac{it^2}{2\phi_P}\right) \int_{-\infty}^{\infty} s(\tau) \exp\left(\frac{i\tau^2}{2\phi_P} - \frac{i\tau t}{\phi_P}\right) d\tau \quad (2.7b)$$

With the above definitions, expanding the convolution operation, we recast (2.6a):

$$|I(\omega)|^2 \propto \left| \exp\left(-\frac{i\omega^2 \phi_P}{2}\right) \int_{-\infty}^{\infty} S(x) \exp\left(-\frac{ix^2 \phi_P}{2}\right) \exp(ix\omega\phi_P) \frac{dx}{2\pi} \right|^2 = \\ = \left| \int_{-\infty}^{\infty} F_s(x) \exp(ix\omega\phi_P) \frac{dx}{2\pi} \right|^2 \propto |f_s(-\omega\phi_P)|^2 \quad (2.8)$$

The idler is then proportional to the Fresnel Integral of the signal $f_s(t)$ defined in the equivalent temporal coordinate:

$$t_{eq} = -\omega\phi_P \quad (2.9)$$

Considering the equivalent definitions for the Fresnel Integral of the PUT:

$$F_E(\omega) = E(\omega) \exp\left(-\frac{i\omega^2 \phi_P}{2}\right) \quad (2.10a)$$

$$f_e(t) = \frac{1}{\sqrt{2i\pi\phi_P}} \exp\left(\frac{it^2}{2\phi_P}\right) \int_{-\infty}^{\infty} e(\tau) \exp\left(\frac{i\tau^2}{2\phi_P} - \frac{i\tau t}{\phi_P}\right) d\tau \quad (2.10b)$$

we have:

$$|I(\omega)|^2 \propto |f_s(t_{eq})|^2 = \left| f_e\left(t_{eq} + \frac{\Delta t}{2}\right) + f_e\left(t_{eq} - \frac{\Delta t}{2}\right) \right|^2 \quad (2.11a)$$

This identity is the core of the FLEA algorithm, which relates the idler to the signal with the only constraints *on the pump pulse* being those given by 2.5a-2.5b. Eq.(2.11a) contains the information on the phase $\varphi_{fe}(t_{eq})$ of $f_e(t_{eq})$, better seen by expressing (2.11a) as the sum of the three terms:

$$|I(\omega)|^2 \propto dc(t_{eq}) + ac_+(t_{eq}) + ac_-(t_{eq}) \quad (2.11b)$$

defined by the following relations:

$$dc(t_{eq}) = \left| f_e\left(t_{eq} + \frac{\Delta t}{2}\right) \right|^2 + \left| f_e\left(t_{eq} - \frac{\Delta t}{2}\right) \right|^2 \quad (2.12a)$$

$$ac_\pm(t_{eq}) = \left| f_e\left(t_{eq} + \frac{\Delta t}{2}\right) \right| \left| f_e\left(t_{eq} - \frac{\Delta t}{2}\right) \right| \exp\left[\pm i\Delta\theta(t_{eq})\right] \quad (2.12b)$$

where $\Delta\theta(t_{eq})$ is the differential phase of the Fresnel Integral of the PUT:

$$\Delta\theta(t_{eq}) = \varphi_{fe}\left(t_{eq} + \frac{\Delta t}{2}\right) - \varphi_{fe}\left(t_{eq} - \frac{\Delta t}{2}\right) \quad (2.12c)$$

Eq.(2.12a) is the sum of the spectral moduli of the Fresnel Integral of the relatively delayed PUTs and does not contain any information on the phase of the PUT. We refer to this component as a "direct current" (DC) component as it is centered at zero in the equivalent frequency domain $\omega_{eq}$. The phase information (2.12c) is contained in Eq.(2.12b) and, as discussed later, its phase contribution typically result in a frequency displacement in the transformed domain. For this reason, we refer to these terms as "alternate current" (AC) terms. In a large number of cases the AC terms can be isolated from the DC components using a filtering procedure, and then be transformed back to extract the differential phase 2.12c. This is the very robust procedure introduced by Takeda, Ina and Kobayashy [15], generally employed in the classical SPIDER approach to extract the phase information of the optical field from the measured optical interferogram.

## 2.1 High pump chirp approximation and standard X-SPIDER relation

Before presenting further details on the FLEA, we show that the standard relationships used in the X-SPIDER phase reconstruction process can be obtained from our more general expressions in (2.11) and (2.12) by carrying out a Fraunhofer approximation of the Fresnel integral of the PUT, to explicitly obtain the shape of the PUT spectrum:

$$f_e(t_{eq}) \propto e(t_{eq}) * \exp\left(\frac{it_{eq}^2}{2\phi_P}\right) \approx E\left(-\frac{t_{eq}}{\phi_P}\right) \exp\left(\frac{it_{eq}^2}{2\phi_P}\right) \quad (2.13a)$$

This approximation is strictly valid when:

$$\Delta \tau_E^2 \ll 2\pi \phi_P \approx 2\pi \frac{\Delta \tau_P}{\Delta \omega_P} \qquad (2.13b)$$

When the pump spectrum is larger or equal to the PUT spectrum, as usually verified in XSPIDER set-ups, 2.13b necessarily implies that $\Delta \tau_E \ll \Delta \tau_P \approx \Delta \omega_P \phi_P$, i.e. the PUT temporal window must be much shorter than the pump time duration. The Fresnel Integral is thus proportional to the spectrum of the PUT. Substituting (2.13a) into (2.11a) and considering the definition of the temporal coordinate 2.9, we derive the classical expression for the spectral-sheared interference pattern [1]:

$$|I(\omega)|^2 \propto \left| E\left(\omega + \frac{\Omega}{2}\right) \exp\left(+i\omega \frac{\Delta t}{2}\right) + E\left(\omega - \frac{\Omega}{2}\right) \exp\left(-i\omega \frac{\Delta t}{2}\right) \right|^2 \qquad (2.14a)$$

The pattern contains two spectral replicas of the PUT shifted in time and in frequency by $\Delta t$ and $\Omega$, respectively, the latter being the spectral shear defined as:

$$\Omega = \frac{\Delta t}{\phi_P} \qquad (2.14b)$$

The DC and AC components of Eq.(2.12) can then be written as:

$$dc(\omega) = \left| E\left(\omega + \frac{\Omega}{2}\right) \right|^2 + \left| E\left(\omega - \frac{\Omega}{2}\right) \right|^2 \qquad (2.15a)$$

and:

$$ac_\pm(\omega) = \left| E\left(\omega + \frac{\Omega}{2}\right) \right| \left| E\left(\omega - \frac{\Omega}{2}\right) \right| \exp\left[\pm i\Delta\theta(\omega)\right] \qquad (2.15b)$$

Te latter is a measurable quantity that depends explicitly on the spectral differential phase of the PUT:

$$\Delta\theta(\omega) = \Delta\varphi_E(\omega) + \omega\Delta t = \varphi_E\left(\omega + \frac{\Omega}{2}\right) - \varphi_E\left(\omega - \frac{\Omega}{2}\right) + \omega\Delta t \qquad (2.15c)$$

As anticipated, the derived equations are the classical expressions exploited by the standard SPIDER method, which directly targets the reconstruction of the spectral phase of the PUT.

## 3. Phase recovery algorithms

We next discuss the FLEA to reconstruct the PUT, by describing step by step the procedure in comparison with the classic approach. Similarly to the conventional phase-reconstruction SPIDER algorithm, FLEA requires a-priori knowledge on the time delay $\Delta t$ and the effective chirp of the interaction $\phi_P$.

*3.1 Step1: Takeda procedure for the extraction of the differential phase from the interferogram.*

The Takeda procedure [1,13-15] can be in general applied to interferograms of different nature. The interferogram 2.11a (or 2.14a in its approximated form) can be described as the superposition of the "DC" and "AC" contributions in equations 2.12 (a-b). The AC contributions are responsible for the interference fringes and as a result, they can be extracted from the interferogram by applying a Fourier Transform procedure. Specifically, the Fourier Transform of the DC and AC components defined in Equations 2.12a-c in the space of $\omega_{eq}$ are:

$$DC(\omega_{eq}) = \cos\left(\frac{\Delta t \omega_{eq}}{2}\right) \int_{-\infty}^{\infty} F_e\left(\frac{s+\omega_{eq}}{2}\right) F_e^*\left(\frac{s-\omega_{eq}}{2}\right) \frac{ds}{2\pi} \quad (3.1a)$$

and

$$AC_{\pm}(\omega_{eq}) = \frac{1}{2} \int_{-\infty}^{\infty} F_e\left(\frac{s+\omega_{eq}}{2}\right) F_e^*\left(\frac{s-\omega_{eq}}{2}\right) \exp\left(\mp \frac{is\Delta t}{2}\right) \frac{ds}{2\pi} \quad (3.1b)$$

Roughly, if the fringes are very dense as compared with the PUT spectral bandwidth, the AC contributions will be well separated from the DC contribution in the transformed space, as sketched in Fig. 2. Since the AC components contain information on the spectral phase, one of them is filtered and then transformed back. The phase of the resulting signal is the differential spectral phase $\Delta\theta(t_{eq})$. It must be recalled that this operation cannot give any information on the *sign* of the differential phase, hence there is an ambiguity, rigorously the quantity obtained at this step is:

$$\mu\Delta\theta(t_{eq}) \quad (3.1.c)$$

Where $\mu$ can be $\pm 1$ and is not known.

In general, the issue of the separability of the terms in Equations (3.1a) and (3.1b) in the global Fourier Transform of the interferogram is nontrivial and depends on the unknown phase of the PUT. In Fig. 3 we report the DC and AC terms in the 2D phase space, i.e. as a function of the frequency $\omega_{eq}$ and delay $\Delta t$ variables, for a Gaussian pulse with first order and second order chirp. In the space $(\omega_{eq}, \Delta t)$, we note that the AC terms have the shape of Wigner functions [1,23-24] - specifically, they are the cross-Wigner function of the Fresnel Integral of the PUT $f_e(t)$ and its time reversed copy $f_e(-t)$ in $(\omega_{eq}, \Delta t)$ and $(\omega_{eq}, -\Delta t)$, respectively. In these types of functions, a first order chirp generates a shear of the function along the straight line $\omega_{eq} = \Delta t/\phi$ where $\phi$ is the *total* first order chirp of the Fresnel integral of the PUT. For large values of $\phi$ the spectral content of the AC functions is nearly entirely localized along the loci $\omega_{eq} = \pm \Delta t/\phi$; in the same way as the spectral content of the DC component shrinks around $\omega_{eq} = 0$. This means that, roughly speaking, if the phase of $F_e(\omega_{eq})$ has a dominant first order chirp (as it *always* occurs in the classical SPIDER approximation, which is dominated by the pump chirp $\phi_P$) encompassing both the pump and the PUT chirps, the spectral separation among the DC and AC terms is always ensured for a certain time delay $\Delta t$.

Although the quantities are interpreted differently, Step1 is the same as that conventionally used in the classical SPIDER reconstruction algorithm: the phase $\Delta\theta$ used in the classical algorithm is indeed the same quantity extracted here (i.e. function given by (2.12c)), but interpreted in its approximated form (2.15c).

*3.2 Step2: extraction of the phase $\varphi_{fe}(t_{eq})$ from $\Delta\theta(t_{eq})$.*

The phase $\Delta\theta(t_{eq})$ must be processed to obtain the phase of the Fresnel Integral $f_e(t_{eq})$. It is important to note that although the phase extraction cannot provide the sign of $\Delta\theta$, this information can be recovered with some strategies described below. For a sufficiently small temporal delay $\Delta t$, the differential phase is the derivative of the phase $\varphi_{fe}(t_{eq})$:

$$\mu\Delta\theta(t_{eq}) = \mu\left(\varphi_{fe}\left(t_{eq}+\frac{\Delta t}{2}\right)-\varphi_{fe}\left(t_{eq}-\frac{\Delta t}{2}\right)\right) \approx \mu\frac{d\varphi_{fe}(t_{eq})}{dt_{eq}}\Delta t \quad (3.2)$$

We can directly integrate the result of Step 1 using the novel temporal coordinates introduced in Eq.(2.9):

$$\mu\int\Delta\theta(t_{eq})\frac{dt_{eq}}{\Delta t} = -\mu\int\Delta\theta(\omega)\frac{\phi_P d\omega}{\Delta t} \approx \mu\varphi_{fe}(t_{eq}) \quad (3.3)$$

In the last equality we took into account the physical quantities. It is interesting to notice that even in this Step 2, FLEA follows closely the standard extraction procedure. If the approximated relation for the differential phase 2.15c holds, taking into account 2.14b, we have:

$$-\mu\int\Delta\theta(\omega)\frac{\phi_P d\omega}{\Delta t} = -\mu\int\Delta\varphi_E(\omega)-\omega\Delta t\frac{d\omega}{\Omega} \approx \mu\left(\varphi_E(\omega)-\frac{\omega^2\phi_P}{2}\right) \quad (3.4)$$

Eq.(3.4) directly relates the differential phase obtained from Step 1 with the spectral phase of the PUT $\varphi_E(\omega)$. The additional quadratic term can be easily removed as the pump effective chirp is known (indeed, the linear term $\omega\Delta t/\Omega$ is usually removed before the integration). As the chirp of the pump is supposed dominant over the PUT phase for the classical SPIDER approximation, the ambiguity on the sign of $\Delta\theta$ is removed in this case.

For large spectral shears, the derivative approximation is no more valid and several effective approaches have been proposed to retrieve the phase of the spectral PUT. Among them, a very popular concatenation method reconstructs exactly the phase even for large spectral shears. This method is discussed in detail elsewhere [1,14], including discussions on the accuracy of higher order integration and concatenation based techniques for different pulse profiles [25-26]. It is beyond the scope of this paper to discuss in detail how these methods can improve the performances for the FLEA. However, as Step 2 follows exactly the standard reconstruction procedure, concatenation and higher order integration techniques can be equivalently employed also for the FLEA as done in the classical algorithm.

*3.3 Step 3: Reconstruction of the full function*

The amplitude of a single idler replica is measured independently and associated with the reconstructed phase. In the standard SPIDER approach, this quantity represents the PUT spectrum: the measurement is then complete. The temporal profile can be obtained with a simple Fourier Transformation. It is important to note that the SPIDER method does not have any ambiguity in the sign of the temporal axis because the sign of the phase is determined as discussed previously. If the spectrum of the PUT is known, it can be associated with the reconstructed phase for better accuracy.

In the case of the FLEA, the spectral measurement of a single replica of the idler associated to the reconstructed phase is interpreted as the Fresnel Integral a of the PUT $f_e(t_{eq})$, which then needs to be inverted to obtain the PUT. As noted previously, the phase extraction procedure does not provide information on the sign of the phase. Strictly speaking, we have to take into account the two relations:

$$\left|f_e(t_{eq})\right|\exp\left(i\varphi_{f_e}(t_{eq})\right) = f_e(t_{eq}) \quad (3.5a)$$

$$\left|f_e(t_{eq})\right|\exp\left(-i\varphi_{f_e}(t_{eq})\right) = f_e^*(t_{eq}) \quad (3.5b)$$

Eq.(3.5a) is the correct Fresnel integral of the PUT, while Eq.(3.5b) is its conjugate copy, leading to a *wrong result*. To obtain the PUT we Fourier Transform them both. We then obtain the correct Fresnel Integral of the PUT in frequency (Eq.(3.6a)) and its conjugate and *frequency reversed* copy (3.6b):

$$F_e(\omega_{eq}) = \left|E(\omega_{eq})\right|\exp\left(i\varphi_E(\omega_{eq}) - i\frac{\phi_P \omega_{eq}^2}{2}\right) \quad (3.6a)$$

$$F_e^*(-\omega_{eq}) = \left|E(-\omega_{eq})\right|\exp\left(-i\varphi_E(-\omega_{eq}) + i\frac{\phi_P \omega_{eq}^2}{2}\right) \quad (3.6b)$$

where we use Eq.(2.9a). To obtain the PUT we subtract the phase $\exp(-i\phi_P \omega_{eq}^2/2)$ and obtain the two results:

$$E(\omega_{eq}) = \left|E(\omega_{eq})\right|\exp\left(i\varphi_E(\omega_{eq})\right) \quad (3.7a)$$

$$E'(\omega_{eq}) = |E(-\omega_{eq})| \exp(-i\varphi_E(\omega_{eq}) + i\phi_P \omega_{eq}^2) \quad (3.7b)$$

The Fourier Transform of the Fresnel Integral Eq.(3.6a) has the same spectral amplitude as the PUT. If the spectrum of the PUT is not symmetric, Equations (3.6b) and (3.7b) representing the incorrect result can easily be recognized by simply comparing them with a spectral measurement of the PUT. Otherwise, for spectrally symmetric waveforms, additional information on the PUT is needed. A second measurement with a different pump chirp will get rid of this ambiguity; however, very often it is simply possible to infer the correct phase as the two results differ by a large phase term. As in the case of standard SPIDER, the correct result given by Eq.(3.7a) does not have any ambiguity in the time domain.

## 4. Numerical Examples

Summarizing the preceding discussion, the FLEA removes the high chirp approximation for the pump pulse and is subject to the following main limitations:

- The time duration of the pump (under approximation Eq.(2.5)), which defines the temporal window of the interaction, must be known a priori.
- The applicability of the Takeda phase extraction procedure. This depends on the phase of the unknown PUT and is not known a priori. In general, a larger pump chirp guarantees better performance.
- It is ambiguous in the sign of the phase for pulses with symmetric spectral amplitude. However this can simply be solved via an additional measurement.

To better quantify the limitations imposed by the first two points, and also to compare with the standard extraction procedure, we calculate the accuracy of FLEA on benchmark test pulses[25-26]. Specifically, we consider a pulse with Gaussian spectral amplitude:

$$E(\omega) = \exp\left(-\frac{\omega^2}{\Delta\omega_E^2}\right) \qquad (4.1)$$

We use this spectrum with different spectral phase profiles, such as phase jumps, parabolic, cubic and quartic phases. This benchmark pulse has already been significantly discussed in connection with calculations of the accuracy and precision of SPIDER reconstruction techniques [25-26].

The pump pulse is assumed to possess a nearly flat-top spectral amplitude. Specifically, it is assumed to have a super-Gaussian shape and parabolic phase:

$$P(\omega) = \exp\left(-\frac{\omega^{16}}{\Delta\omega_P^{16}}\right)\exp\left(-\frac{i\omega^2 \phi_P}{2}\right) \qquad (4.2)$$

In the numerical tests we use pulses with frequency bandwidths of the order of those expected in ultrafast optical communications, i.e. $\Delta\omega_E=2\pi\times 1$THz, $\Delta\omega_P=2\pi\times 2$THz, and $\phi_P$ with values varying from 10ps$^2$ - i.e. the pump covers a temporal window of approximately 250ps (full width) - to 20ps$^2$ - i.e. covering a 500ps time window. We discuss the performance of the method both for TWM (SFG) and DFWM cases. As discussed above, since we employ an almost flat-top pump, the effective temporal window in the two cases is the same for the same pump, while the effective chirp is opposite and half of the pump chirp for the DFWM case.

We used $2^{16}$ samples with a temporal resolution of 40fs. To simulate experimental conditions, we also took into account the limited bandwidth of the spectrometer, which we set to 1GHz. This translates to an equivalent temporal window of 600ps for the measured spectrogram, imposing a weaker temporal restriction than the pump.

To quantify the accuracy, we evaluated the RMS error $\varepsilon$ of the reconstructed PUT $E_R(\omega)$, compared with the original waveform $E(\omega)$:

$$\varepsilon = \left[\frac{\int_{-\infty}^{\infty}|E_R(\omega)-E(\omega)|^2 \frac{d\omega}{2\pi}}{\int_{-\infty}^{\infty}|E(\omega)|^2 \frac{d\omega}{2\pi}}\right]^{1/2} \qquad (4.3)$$

Usually, RMS values below 0.02 are considered to be excellent reconstructions, while values above 0.2 are considered to fail the reconstruction [25-26].The delay between the two PUT replicas is set to 6.5ps, and is chosen to minimize the error introduced by the integration of the differential phase, while allowing the filtering procedure for the standard algorithm.

*4.1 Phase-jump pulses.*

We next apply phase jumps given by:

$$\varphi_E(\omega) = \frac{\pi}{2}\tanh\left(\alpha\frac{\omega}{\Delta\omega_E}\right) \quad (4.4)$$

Where $\alpha$ is a parameter that indicates the steepness of the phase jump. Fig. 4 shows the spectral amplitude (Fig. 4 a, black solid line) and phase for $\alpha=10$ and $\alpha=40$ (Fig. 4 b, black solid line and Fig. 4 d dashed lines, respectively). In Fig. 4(a) and (c) we show the intensity of the idler replica generated by a SFG nonlinear interaction, for $\alpha=10$ and $\alpha=40$ (red solid and dashed lines, respectively). Here the pump possesses $\phi_P =10ps^2$, $\Delta\tau_P =250ps$. This in turn translates into a significant distortion of the spectrum of the idler replicas when compared with the original PUT spectrum. As discussed above, the distortion of the idler spectrum for the single idler replica is a clear indication that the high pump chirp approximation is exceeded.

The reconstructed phases using both the classical algorithm and the FLEA are shown in Fig. 4, in red and green, respectively. (a-b) report the PDS for $\alpha=10$ and $\alpha=40$, respectively. The green plots indicate the PSD obtained with the FLEA reconstruction: remarkably the algorithm largely mitigates the distortion on the spectral amplitude visible in the idler replica reported in red. The reconstructed phase is in Fig. (c-d-e). For $\alpha=10$ both algorithms show good accuracy while for $\alpha=40$ the FLEA shows better performance.

In Fig. 5 (a) and (b) the RMS error as a function of $\alpha$ is reported for both the standard algorithm (red) and the FLEA (green). Fig. 5(a) shows the error obtained with SFG. In Fig. 5(b) we used DFWM. We report in dark red and dark green the reconstruction obtained with a pump chirp $\phi_P =10ps^2$ while light green is used for the case for a pump chirp $\phi_P =20ps^2$. In both cases, FLEA shows a significant reduction of the RMS error for high values of the phase jump parameter α, associated to longer temporal pulses. It is interesting to notice that the RMS error for the SFG case $\phi_P$ =10ps² follows closely the RMS error for the case DFWM $\phi_P$ =20ps². As we remarked above, the pump chirp counts half in the case of a DFMW interaction with respect to a TWM case. Hence, in the two cases the nonlinear interaction provides the same magnitude of equivalent chirp. Conversely, the DFWM interaction for $\phi_P$ =20ps² possesses an equivalent temporal window that is approximately twice the case of SFG with $\phi_P$ =10ps², and approximately equivalent to the case of SFG with $\phi_P$ =20ps².

Indeed, the error is lower for larger pump chirps while it is insensitive to the dimension of the time window of the interaction, witnessing that the error for large α is not associated to the time window of the pump, but can be related to the increasing of overlapping of the AC spectral components (3.3b) for large α, which ultimately limits the phase extraction procedure. In any case, the error remains remarkably low even for extremely steep phase jumps.

*4.2 High order phase dispersion.*

Here we tested the performances of the FLEA applying first, second and third order dispersion to the PUT discussed above. The results in Fig. 6(a-d) are relative to first order dispersion phases for classic algorithm and FLEA for a SFG (a-b) and for classic algorithm and FLEA for DFWM (c-d). They evidence a greatly extended range of consistency for the FLEA algorithm. The green dotted lines in (b) mark the maximum effective chirp that can be addressed due to the temporal limitation imposed by the pump temporal window. These lines are the same in the case of DFMW in (d). The cyan dotted lines in (b) report a "blind region" of the algorithm: for these values the PUT chirp exactly compensates for the pump chirp. In this case the overall first order chirp of the Fresnel Integral is zero. As discussed in paragraph 3.2, the fringes are not formed in the interferogram and the Takeda Extraction procedure cannot be applied. The same region in (d) for the DFWM case is formed around the chirp values that compensate half of the pump chirp, with opposite sign.

Fig. 7 shows the results for predominantly second order chirped pulses, with the same convention as of Fig. 6. The FLEA also extends the working regime of the X-SPIDER device in this case. The limitation for large third order chirp arises from the Takeda phase extraction procedure for the set of parameters exploited here, for pulses with both large positive and negative second order chirp, in a similar fashion to the example reported in Fig. 3. For this reason, the DFWM reconstruction for the FLEA shows a narrower range of consistency than the SFG case when the same pump is used, as the effective chirp of the nonlinear interaction is half of the pump chirp. However, in both cases the FLEA significantly extends the performance of the X-SPIDER setup.

Figure 8 shows the results for predominantly third order chirped pulses. The limitation for large third order chirp also arises from the Takeda phase extraction procedure for negative chirp, while the temporal window of the pump limits the reconstruction of the PUT with positive third order dispersions.

## 5. Conclusions

We have discussed the performances of an innovative extraction algorithm for X-SPIDER , that we term FLEA. We have discussed in detail the FLEA approach, and defined its limits for PUT reconstruction in terms of the physical parameters of the pump and the PUT. We evaluated numerically the performance of the FLEA, also in comparison with the standard X-SPIDER algorithm, for a conventional, comprehensive set of bench pulses: in all the addressed cases FLEA demonstrates a remarkably larger range of consistency in the reconstruction, evidencing a significant improvement in terms of time-bandwidth product of the reconstructed pulses.

**Acknowledgments**

This work was supported by the Natural Sciences and Engineering Research Council and by the Australian Research Council Discovery and Centers of Excellence Research Program.

1. I. A. Walmsley and C. Dorrer, "Characterization of ultrashort electromagnetic pulses", Adv. Opt. Photon. **1**, 308-437 (2009).
2. R. Slavik, et al., "All-optical phase and amplitude regenerator for next-generation telecommunications systems", Nature Photon. **4**, 690-695 (2010).
3. R. J. Essiambreet al., "Capacity Limits of Optical Fiber Networks", J. Lightwave Technol. **28**, 662-701 (2010).
4. N. K. Fontaine, R. Scott, L. Zhou, F. M. Soares, J. Heritage and S. Yoo, "Real-time full-field arbitrary optical waveform measurement", Nature Photon. **4**, 248-254 (2010).
5. Z. Jiang, C.-B. Huang, D. E. Leaird and A. Weiner, "Optical arbitrary waveform processing of more than 100 spectral comb lines", Nature Photon. **1**, 463-467 (2007).
6. S. T. Cundiff and A. M. Weiner, "Optical arbitrary waveform generation", Nature Photon. **4**, 760-766 (2010).
7. M. A. Foster et al., "Silicon-chip-based ultrafast optical oscilloscope," Nature **456**, 81-84 (2008).
8. R. Salem et al., "Optical time lens based on four-wave mixing on a silicon chip," Opt. Lett. **33**, 1047-1049 (2008).
9. A. Pasquazi et al., "Time-Lens Measurement of Sub-picosecond Optical Pulses in CMOS Compatible High-Index Glass Waveguides," J. Sel. Top. Quantum Electron **18**, 629-636 (2012).
10. E. K. Tien, X. Z. Sang, F. Qing, Q. Song and O. Boyraz, "Ultrafast pulse characterization using cross phase modulation in silicon," Appl. Phys. Lett. **95**, 051101 (2009).
11. C. V. Bennett, R. P. Scott and B. H. Kolner, "Temporal magnification and reversal of 100 Gb/s optical-data with an up-conversion time microscope," Appl. Phys. Lett. **65**, 2513-2515 (1994).
12. R. Trebino, "Frequency Resolved Optical Gating: the Measurement of Ultrashort Optical Pulses," (Kluwer Academic, 2002).
13. C. Iaconis and I. A. Walmsley, "Spectral phase interferometry for direct electric-field reconstruction of ultrashort optical pulses," Opt. Lett. **23**, 792-794 (1998).
14. C. Iaconis and I. A. Walmsley, "Self-referencing spectral interferometry for measuring ultrashort optical pulses," IEEE J. Quant. Electron. **35**, 501-509 (1999).
15. M. Takeda, H. Ina and S. Kobayashi, "Fourier-transform method of fringe-pattern analysis for computer-based topography and interferometry," J. Opt. Soc. Am. **72**, 156-160 (1982).
16. M. Hirasawa et al., "Sensitivity improvement of spectral phase interferometry for direct electric-field reconstruction for the characterization of low-intensity femtosecond pulses," Appl. Phys. B **74** (Suppl. S), S225-S229 (2002).
17. P. Londero, M. E. Anderson, C. Radzewicz, C. Iaconis, and I. A. Walmsley, "Measuring ultrafast pulses in the near-ultraviolet using spectral phase interferometry for direct electric field reconstruction," J. Mod. Opt. **50**, 179–184 (2003).
18. L. Gallmann et al., "Characterization of sub-6-fs optical pulses with spectral phase interferometry for direct electric-field reconstruction," Opt. Lett. **24**, 1314-1316 (1999).
19. C. Dorrer et al., "Single-shot real-time characterization of chirped-pulse amplification systems by spectral phase interferometry for direct electric-field reconstruction," Opt. Lett. **24**, 1644-1646 (1999).
20. J. Wemans, G. Figueira, N. Lopes and L. Cardoso, "Self-referencing spectral phase interferometry for direct electric-field reconstruction with chirped pulses," Opt. Lett. **31** (14), 2217-2219 (2006).
21. A. Pasquazi et al., "Sub-picosecond phase-sensitive optical pulse characterization on a chip," Nature Photonics **5**, 618-623 (2011).
22. J. Cohen, P. Bowlan, V. Chauhan, P. Vaughan, and R. Trebino, "Single-shot multiple-delay crossed-beam spectral interferometry for measuring extremely complex pulses," Opt. Comm., **284**, 3785-3794 (2011).
23. L. Cohen, "Wigner distribution for finite duration or band-limited signals and limiting cases," IEEE Transactions on Acoustics, Speech and Signal Processing, **35**, 796–806, (1987).
24. F. Hlawatsch, "Interference terms in the Wigner distribution," Digital Signal Processing **84**, 363–367, (1984).
25. C. Dorrer and I. A. Walmsley, "Precision and consistency criteria in spectral phase interferometry for direct electric-field reconstruction," J. Opt. Soc. Am. B **19**, 1030-1038, (2002).
26. C. Dorrer and I. A. Walmsley, "Accuracy criterion for ultrashort pulse characterization techniques: application to spectral phase interferometry for direct electric field reconstruction," J. Opt. Soc. Am. B **19**, 1019-1029, (2002).

Table 1
Equivalent quantities for the nonlinear interaction in different frequency mixing cases.

| | TWM (SFG) | TWM (DFG) | DFWM |
|---|---|---|---|
| Temporal window $\Delta \tau_{PN}$ (2.5) | Gaussian (waist): $\Delta\omega_P \phi_P$ <br> Flat top (full width): $\Delta\omega_P \phi_P$ | Gaussian (waist): $\Delta\omega_P \phi_P$ <br> Flat top (full width): $\Delta\omega_P \phi_P$ | Gaussian (waist): $\Delta\omega_P \phi_P / \sqrt{2}$ <br> Flat top (full width): $\Delta\omega_P \phi_P$ |
| Equivalent time (2.9) | $t_{eq} = -\omega \phi_P$ | $t_{eq} = \omega \phi_P$ | $t_{eq} = \omega \phi_P / 2$ |
| Fresnel Integral (2.10b) | $f_e(t) \propto \exp\left(\dfrac{it^2}{2\phi_P}\right) * e(t)$ | $f_e(t) \propto \exp\left(\dfrac{-it^2}{2\phi_P}\right) * e(t)$ | $f_e(t) \propto \exp\left(\dfrac{-it^2}{\phi_P}\right) * e(t)$ |
| Effective chirp | $\phi_P$ | $-\phi_P$ | $-\phi_P / 2$ |

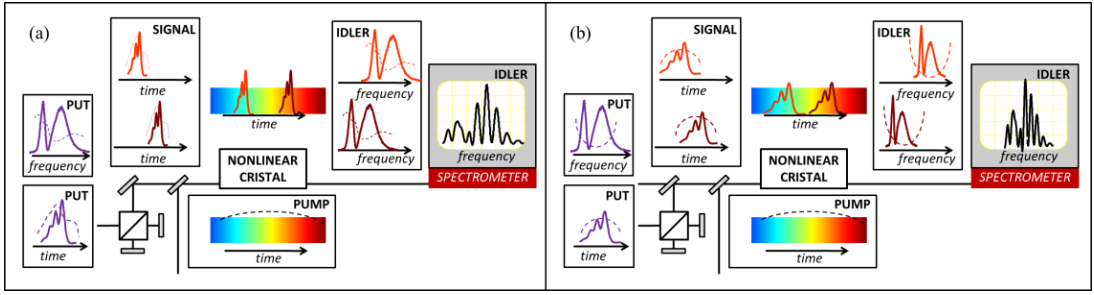

Fig. 1: (a) Sketch of the classical SPIDER set-up for amplitude and phase retrieval. Two delayed replicas of the pulse under test (PUT) interact nonlinearly with a dispersed pump via frequency mixing in a bulk crystal. The idler is composed by two replicas shifted in frequency. Their spectrum is collected with a spectrometer and elaborated to extract the complete information (amplitude and phase) of the PUT. (b) Same interaction with a highly chirped PUT: in this case the two replicas of the signal cover a significant temporal portion of the pump and their spectrum is distorted when compared with the spectrum of the PUT. As a result, the standard algorithm fails to retrieve the PUT. Conversely, the FLEA is able to correctly work on this kind of pulses, by addressing the two replicas of the idler as a Fresnel Integral of the PUT.

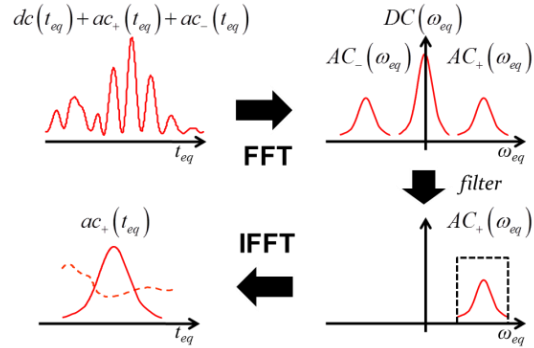

Fig. 2: Phase extraction procedure

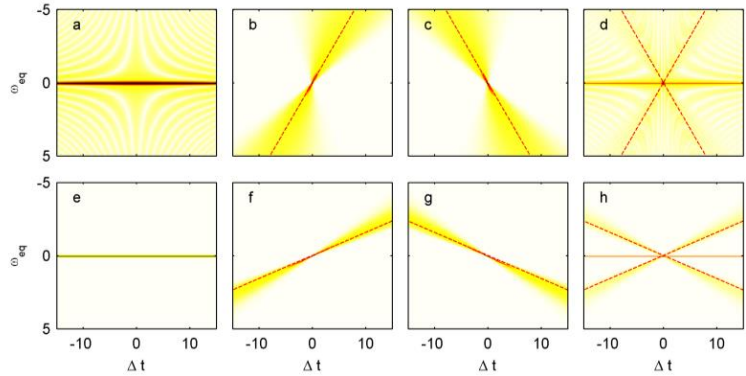

Fig. 3: (a) DC and (b-c) AC components as defined by the 3.3a-b in the phase space ($\omega_{eq}, \Delta t$), for a Gaussian PUT with first and second order chirps. The straight dashed line in red in (b-c) are the loci $\omega_{eq} = \pm \Delta t / \phi$, where $\phi$ is the *total* first order chirp of the Fresnel integral of the PUT, encompassing both the pump and the PUT chirps. In (d) the phase space representation of the complete interferogram is reported: in this case, the separation in frequency is not guaranteed for low values of the time delay $\Delta t$. (e-h) same as (a-d), but with a larger overall total dispersion $\phi$: the spectral content of the DC (e) and AC (f-g) components is localized along $\omega_{eq} = 0$ and $\omega_{eq} = \pm \Delta t / \phi$, respectively: their separation in the interferogram (h) is guaranteed for a large set of delays $\Delta t$.

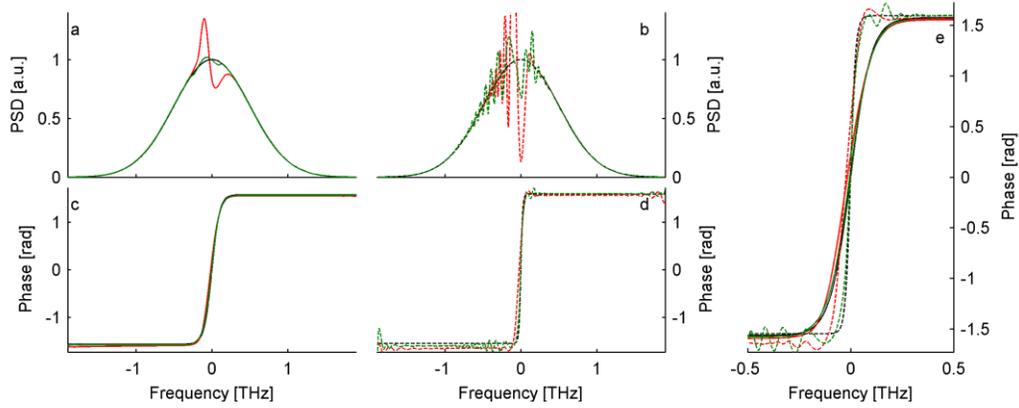

Fig. 4. (a) PSD of the PUT (black) used in the simulation. Idler spectral amplitude for a single replica obtained with a PUT phase jump α=10, and reconstructed PSD with FLEA are displayed in red and green lines, respectively. (b) same as (a) for a PUT phase jump α=40, the different curves are reported using dashed lines in this case. (c) PUT phase for α=10, and reconstructed phases with the classical algorithm and FLEA in black, red and green solid lines, respectively. (d) same as (c) for a PUT phase jump α=40, plots are reported in dashed line in this case. (e) Zoom of the phase plots around the phase jump for the cases above.

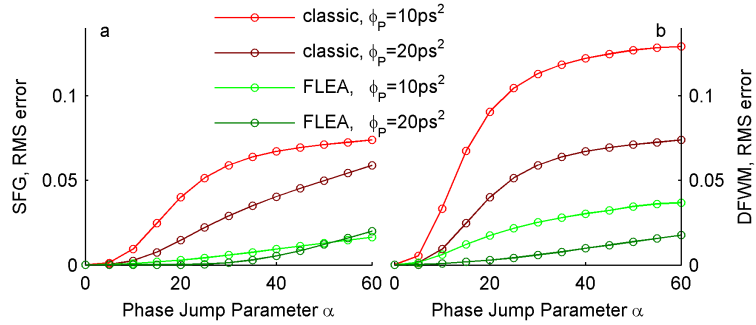

Fig. 5. RMS error as in Eq. (3) for the standard algorithm (red plots) and FLEA (green plots) vs α (higher values are for steeper phase jumps). (a) and (b) show the error calculated by using SFG and DFWM respectively. Light red and light green are for $\phi_P =10\text{ps}^2$, while dark red an dark green are for $\phi_P =20\text{ps}^2$.

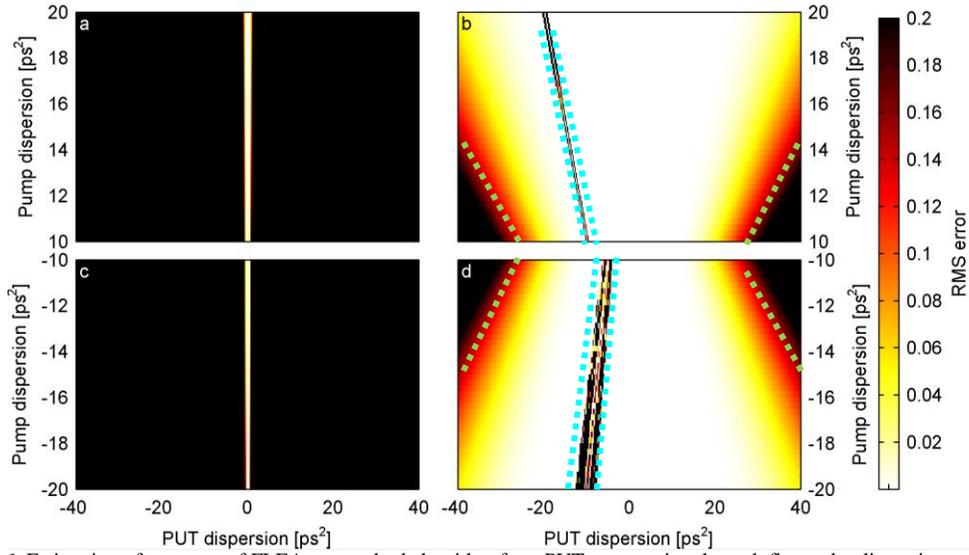

Fig.6. Estimation of accuracy of FLEA vs standard algorithm for a PUT propagating through first order dispersive systems. (a-b) RMS error for the retrieved waveform vs pump (y axis) and PUT(x axis) total dispersions for the SFG case, and (c-d) for the DFWM case; (a-c) standard algorithm (b-d) FLEA. The colorbar reports the RMS error. The picture is saturated for RMS>0.2. In (b) the green dotted line corresponds to the limit imposed by the limitation of the pump temporal window and the cyan dotted lines report the "blind region" of the algorithm, around the opposite of the value of the pump dispersion. This graph clearly evidences that the low-error (white) zone is significantly increased when using FLEA as compared to the standard algorithm. (c-d) same comments as (a-b) for the DFMW case.

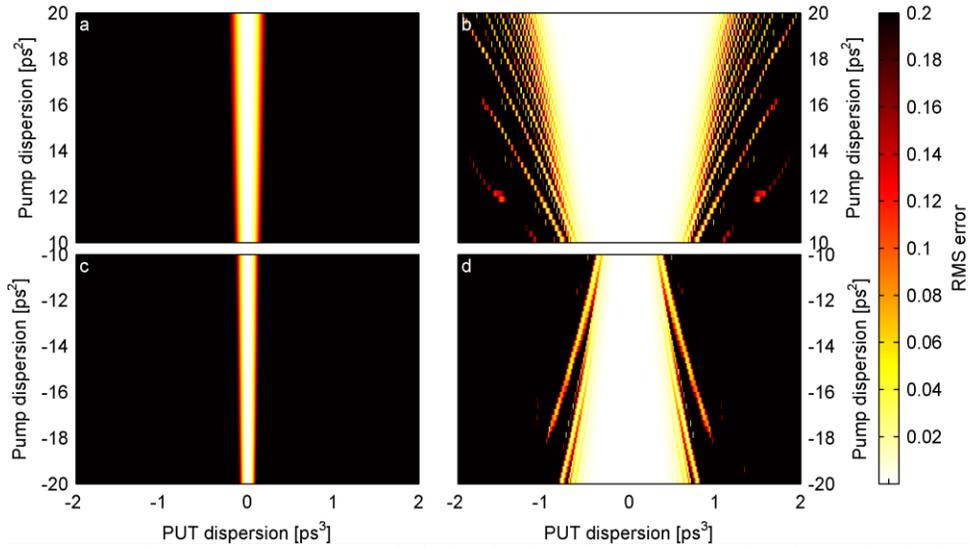

Fig.7. Estimation of accuracy of FLEA vs standard algorithm for a PUT propagating through second order dispersive systems. (a-b) RMS error for the retrieved waveform vs pump (y axis) and PUT(x axis) total dispersions for the SFG case, and (c-d) for the DFWM case; (a-c) standard algorithm (b-d) FLEA. The colorbar reports the RMS error. The picture is saturated for RMS>0.2. This graph clear shows that the low-error (white) zone is significantly increased when using FLEA as compared to the standard algorithm. (c-d) same comments as (a-b) for the DFMW case.

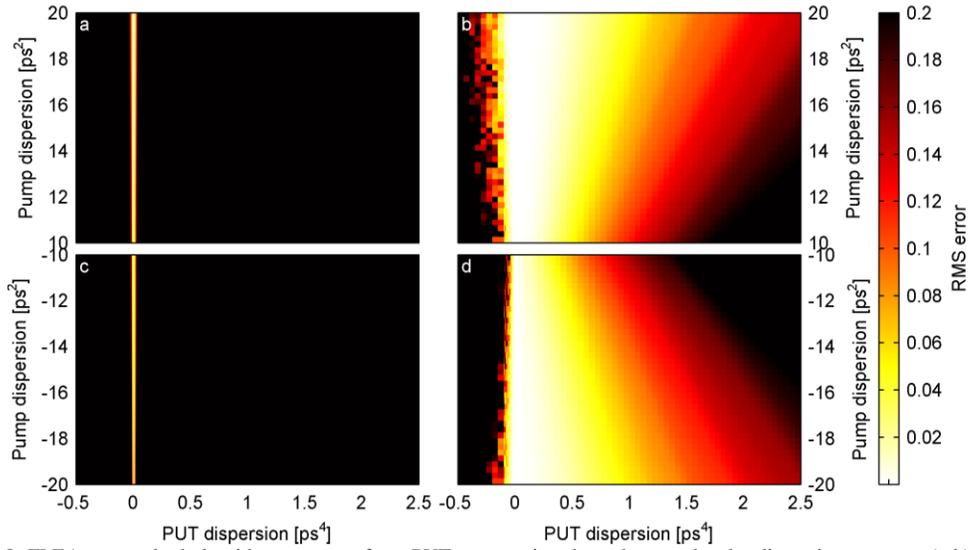

Fig.8. FLEA vs standard algorithm accuracy for a PUT propagating through second order dispersive systems. (a-b) RMS error for the retrieved waveform vs pump (y axis) and PUT(x axis) total dispersions for the SFG case, and (c-d) for the DFWM case; (a) standard algorithm (b) FLEA. The colorbar reports the RMS error. The picture is saturated for RMS>0.2. This graph, it is clear that the low-error (white) zone is significantly increased when using FLEA as compared to the standard algorithm. (c-d) same as (a-b) for the DFMW case.